# LITHIUM NIOBATE ACOUSTIC RESONATORS OPERATING BEYOND 900 °C

*Walter Gubinelli[1], Hasan Karaca[2], Ryan Tetro[1], Sariha N. Azad[2],*
*Philip X.-L. Feng[2], Luca Colombo[1], and Matteo Rinaldi[1]*
[1]Institute for NanoSystems Innovation (NanoSI), Northeastern University, Boston, MA, USA
[2]Department of Electrical & Computer Engineering, University of Florida, Gainesville, FL, USA

## ABSTRACT

In this paper, fundamental shear-horizontal $SH_0$ mode Leaky Surface Acoustic Wave (LSAW) resonators on X-cut lithium niobate leveraging dense and robust electrodes such as gold and tungsten are demonstrated for extreme temperature operation in harsh environments. A novel post-processing approach based on in-band spurious mode tracking is introduced to enable reliable characterization under extreme parasitic loading during testing. Devices exhibit stable performance throughout multiple thermal cycles up to 1000 °C, with an extrapolated electromechanical coupling coefficient $k_t^2 = 25\%$ and loaded quality factor $Q_p = 12$ at 1000 °C for tungsten devices, and $k_t^2 = 17\%$, $Q_p = 100$ at 900 °C for gold devices.

## INTRODUCTION

Microelectromechanical systems (MEMS) are pivotal components in a wide range of high-impact technologies, including timing, sensing, and signal processing [1]. As these systems expand into advanced industrial processing, aerospace, and energy applications, there is an increasing demand for sensing elements capable of maintaining stable and reliable operation under extreme environmental conditions, particularly high temperatures [2] [3] [4]. Traditional silicon-based MEMS technologies are fundamentally limited in such contexts, with electronic components typically failing near 350 °C and mechanical failure typically occurring beyond 500 °C [5]. Similarly, commonly employed metals for MEMS manufacturing, such as aluminum, exhibit poor corrosion resistance and low thermal durability, further restricting their deployment in harsh environments [6].

To overcome these limitations, researchers have explored alternative platforms such as III-nitrides, silicon carbide (SiC), diamond-like carbon, which offer superior mechanical and thermal resilience [7] [8] [9]. SiC particularly stands out for its robustness at temperatures up to 800 °C and its suitability for the manufacturing of high-power electronics.

By a sensing element prospective, Lithium Niobate (LN) resonators for harsh environment sensing have been explored, as this material exhibits high piezoelectric coefficients, excellent chemical stability, and a high Curie temperature (1140 °C) [10].

Furthermore, thin-film transferred lithium niobate on SiC has been demonstrated [11], hinting to the development of fully integrated electronic platform with piezo-on-insulator sensing elements and post-processing high-temperature electronics [12].

This work builds upon earlier findings by extending the operating range of lithium niobate resonators well beyond previously established thermal limits [13]. Fundamental tone shear horizontal $SH_0$ leaky SAW fabricated on bulk X-cut lithium niobate are demonstrated to operate reliably beyond 900 °C, marking the highest verified operation for a microacoustic resonator to date. The devices exhibit high coupling and Q-factor stability under elevated temperatures, enabled by robust electrode configurations and a simplified fabrication process. Despite the inherent challenges associated with high-temperature measurements, promising results are obtained, proving the feasibility of practical chip-level sensing in harsh environments, while addressing the limitations of conventional technologies.

## FABRICATION AND PRELIMINARY CHARACTERIZATION

Building upon previous work [13], fundamental tone shear horizontal $SH_0$ leaky SAW devices are fabricated starting from a bulk X-cut lithium niobate substrate. The electrode materials selected for this study are gold (Au) and tungsten (W). The interdigitated comb-fingers (IDTs) are lithographically patterned with a wavelength (λ) of 6 μm. Shorted lateral reflectors are included to enhance acoustic wave confinement in the active IDT region, thereby improving the quality factor. The fabrication involves two lithographic steps, enabling simultaneous realization of resonators containing tungsten IDTs and combined Au+W contact pads on the same wafer (Fig. 1c). The W electrodes (250 nm thick) are deposited using RF magnetron sputtering, while gold is deposited via electron beam evaporation. The W layer is then etched by ion milling to define both the fingers and pads.

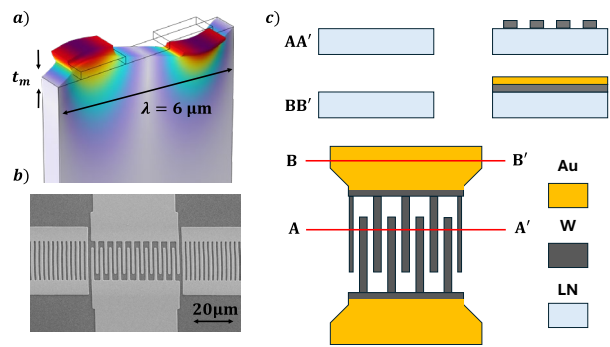

*Figure 1. a) COMSOL® 2.5D simulated mode shape of the fundamental $SH_0$ LSAW at λ = 6 μm; b) SEM image of fabricated gold device; c) Fabrication flow schematic for W LSAW with Au pads.*

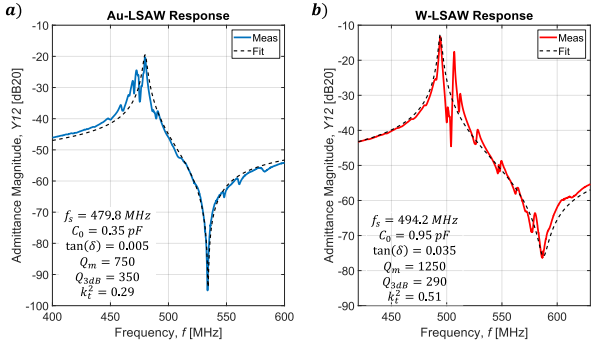

*Figure 2. Measured admittance $Y_{12}$ for (a) gold and (b) tungsten fabricated devices with corresponding mBVD fitting.*

The gold pads are patterned via liftoff on top of the tungsten pads to ease probing and reduce series resistance. Additionally, as shown in Fig. 3b, large pads are added to accommodate the thick probes employed for the high-temperature measurements.

Preliminary characterization is conducted via direct RF Ground-Signal-Ground (GSG) probing coupled with a Vector Network Analyzer (VNA). The measured scattering parameters are converted into admittance $Y_{12}$ parameters and fitted using a single-mode modified Butterworth‑Van Dyke (mBVD) model, as illustrated in Fig. 2. The extracted Figure-of-Merit (FoM) are $FoM = 218$ at 480 MHz and $FoM = 640$ at 494 MHz for the gold and tungsten devices, respectively.

## EXPERIMENTAL METHODS

Due to the unavailability of Ground-Signal-Ground (GSG) RF probes rated for temperatures exceeding 450 °C, frequency measurements at high temperatures were performed using NEXTRON® NXT-1k (MPT-RH) Rhodium DC Probes. The measurements are carried out in a two-port configuration, where two probes are positioned on the large signal pads, while other two additional probes are used to establish a floating ground reference on the chip. The Vector Network Analyzer (VNA) is connected via coaxial cables and the whole heating cell is grounded.

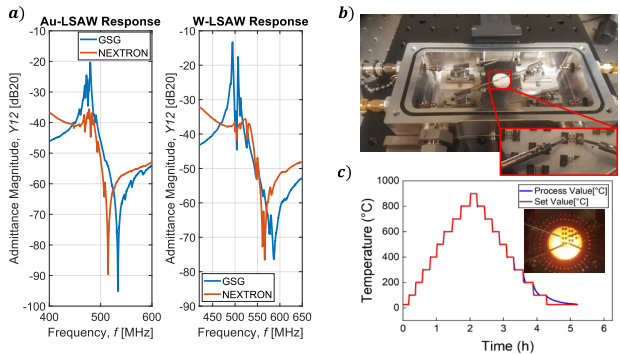

*Figure 3. a) Measured admittance magnitude $Y_{12}$ for Au (left) and W (right) devices, comparing GSG and NEXTRON® probing configurations. The responses were vertically realigned out-of-band to enable spurious mode tracking; b) High temperature vacuum chamber with mounted device; c) Temperature profile used for thermal testing (inset: thermal setup during operation).*

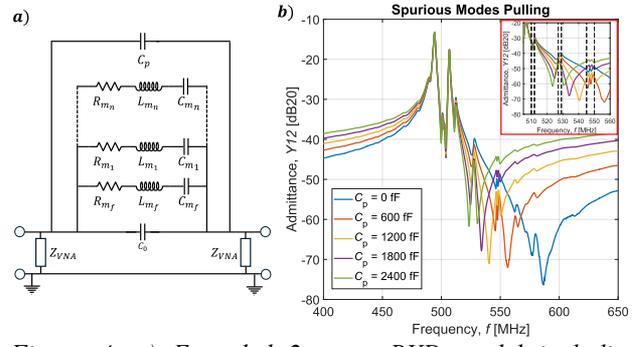

*Figure 4. a) Extended 2-port mBVD model including multiple motional branches and parallel pulling capacitance $C_p$; b) Simulated admittance magnitude $Y_{12}$ showing that spurious modes with lower $k_t^2$ remain relatively unaffected by the capacitive pulling (inset: zoomed-in frequency region with spurious modes highlighted).*

To mitigate oxidation of the probes at elevated temperatures, the measurements are carried out in low vacuum conditions (below 30 mTorr). The temperature is incrementally raised over several hours, maintaining constant intervals to achieve steady-state operation, as depicted in Fig. 3c. The measured S-parameters are then converted to admittance $Y_{12}$.

## METHODOLOGY AND RESULTS
### Radiofrequency Characterization

As depicted in Fig. 3a, the employed probing mechanism introduces complex RF loading effects that significantly hinder the resonators' intrinsic response. Notably, the series resonance frequency $f_s$ appears extensively loaded; concurrently, a noticeable reduction of the apparent electromechanical coupling $k_t^2$ is observed. The setup demonstrates significant sensitivity to variations in the probing configuration; although the measured responses are generally consistent, achieving exact repeatability remains challenging.

Traditional de-embedding methods [14], such as on-chip SOLT calibration, equivalent circuit modeling, and test fixture removal through ABCD-parameters cascading proved inadequate due to the impossibility of de-embedding the DC fixtures employed in the setup. Hence, the tracking of unwanted resonant modes (i.e., in-band spurious modes) is exploited to estimate the evolution of parameters-of-interest (PoI) up to 1000 °C.

As extensively detailed in [15], spurious modes resonances and anti-resonances (poles and zeros) are fairly insensitive to the presence of parasitic reactances due to their limited $k_t^2$ and their intrinsic main mode dependency. In other words, the spurious mode's $f_s$ and $f_p$ can only be marginally shifted by parasitics. For example, Figure 4 illustrates a multi-modal modified Butterworth–Van Dyke (mBVD) model incorporating pulling capacitance in parallel configuration. By simulating the effect of various pulling capacitances on a known device response, it is observed that, although both the parallel resonance quality factor $Q_p$ and electromechanical coupling $k_t^2$ degrade under loaded conditions, spurious mode frequencies remain relatively unaffected.

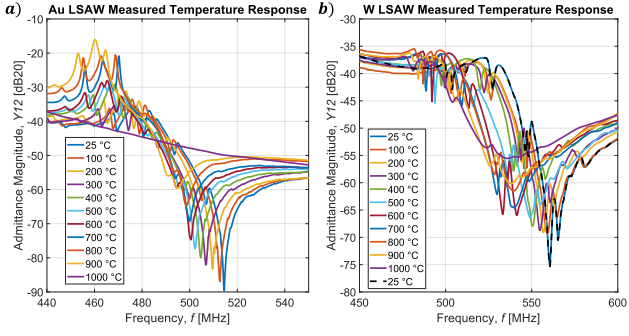

*Figure 5. a) Measured admittance magnitude $Y_{12}$ for Au LSAW devices from 25 °C to 900 °C, showing functional degradation above 800 °C; b) W LSAW devices maintain a consistent response across the full temperature sweep up to 1000 °C and after cooldown.*

The introduction of such parallel parasitic capacitance pushes the series resonant frequency of the spurious mode, as shown in Equation 1. The pulled frequency ($f_{RF}$) is correlated to the undisturbed resonance frequency $f_s$, through a pulling coefficient $\gamma$, defined in Equation 2 [15][16].

$$f_{RF} = f_s \sqrt{\frac{2-\gamma^2}{\gamma^2}} \quad (1)$$

$$\gamma = \sqrt{\frac{C_0 + C_p}{C_0 + C_p + C_m}} \quad (2)$$

Since $\gamma$ is small for a small spurious mode's motional capacitance, $f_{RF}$ will closely track $f_s$. This phenomenon can be used to systematically identify the position of spurious modes, thus enabling device characterization and PoI estimation even under these peculiar high-temperature probing conditions.

**Experimental Results**

Figure 5 shows the extracted admittance parameters for both the gold and tungsten devices for each step in the temperature ramp. Using the method described in the previous section, starting from the converted admittance parameters, the following values (summarized in Table I) are extracted: series resonance frequency $f_s$, parallel resonance frequency $f_p$, apparent electromechanical coupling $k_t^2$, apparent loaded quality factor at antiresonance $Q_p$. The latter can be accurately inferred especially at larger temperatures, when the large impedance falls well above the noise floor of the VNA. The electromechanical coupling and quality factor are evaluated as follows:

$$k_t^2 = \frac{\pi^2}{8}\left(1 - \frac{f_s^2}{f_p^2}\right) \quad (3)$$

$$Q_p = \frac{f_p}{\Delta f_{+3dB}} \quad (4)$$

Additionally, as the identification of the antiresonance is the most straightforward, the first and second order temperature coefficients of frequencies are extracted as follows:

$$\frac{\Delta f_p}{f_{p_0}} = TCF_1 \cdot \Delta T + TCF_2 \cdot \Delta T^2 \quad (5)$$

The extracted values, especially for the first order, are consistent with those previously reported [14] [17]. Au devices show consistent performance up to 900 °C and break down at 1000 °C. On the contrary, W LSAWs survive multiple thermal cycles up to 1000 °C.

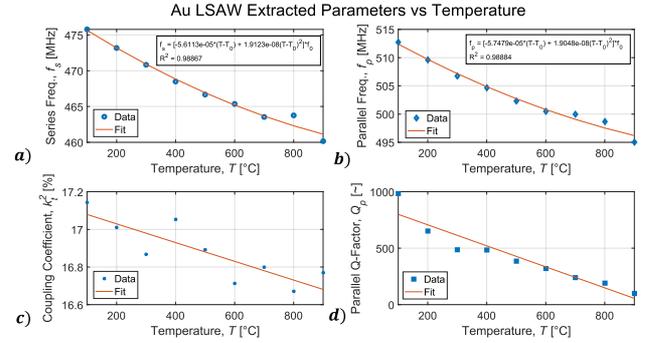

*Figure 6. Extracted temperature-dependent parameters for Au LSAW device up to 900 °C, including (a) series resonance frequency $f_s$, (b) parallel resonance frequency $f_p$, (c) loaded electromechanical coupling $k_t^2$, and (d) parallel quality factor $Q_p$.*

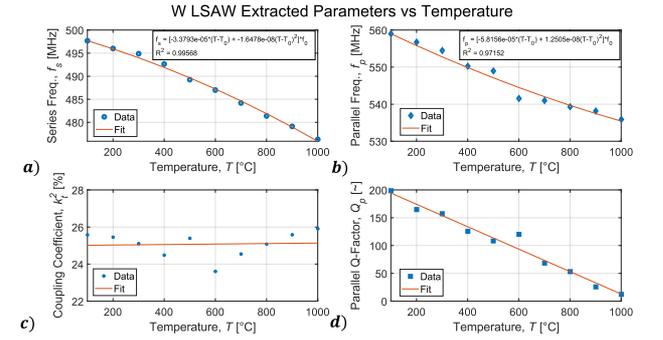

*Figure 7. Extracted temperature-dependent parameters for W LSAW device up to 1000 °C, including (a) series resonance frequency $f_s$, (b) parallel resonance frequency $f_p$, (c) loaded electromechanical coupling $k_t^2$, and (d) parallel quality factor $Q_p$.*

*Table 1. Summary of extracted parameters from high temperature measurements.*

| ~ | **Au LSAW** | **W LSAW** |
|---|---|---|
| $TCF_1$ at $f_p$ | $-56\ ppm/K$ | $-58\ ppm/K$ |
| $TCF_2$ at $f_p$ | $19\ ppb/K^2$ | $12\ ppb/K^2$ |
| $k_t^2$ | $\approx 17\%$ | $\approx 25\%$ |
| $Q_p$ at 25 °C | 950 | 200 |
| $Q_p$ at $T_{max}$ | 100 (900°C) | 12 (1000°C) |

The failure of the gold devices could be traced back to the physical transformation of the metal due to phenomena such as dynamic recrystallization and electromigration that occur at such high temperatures, permanently damaging the electrodes even upon cooling down. Both devices show approximately constant coupling across the full range of temperatures.

## CONCLUSIONS AND REMARKS

X-cut $SH_0$ LSAW resonators on bulk lithium niobate have been demonstrated with stable performance and high coupling up to 1000 °C. Despite parasitic effects introduced by high temperature probing, the proposed analytical method enabled reliable estimation of key parameters. To the best of the authors' knowledge, this marks the first demonstration of chip-scale acoustic resonators enduring repeated thermal cycling at such temperatures. These findings highlight the potential for robust microsystems in extreme environments, although integration with suitable high-temperature electronics, such as GaN or SiC, remains a critical and open challenge.

## ACKNOWLEDGMENTS

The authors would like to thank Northeastern Kostas Cleanroom and Harvard CNS staff.

## CONTACT

Walter Gubinelli, gubinelli.w@northeastern.edu